\documentclass{aastex}
\usepackage{emulateapj5}

\slugcomment{}

\shorttitle{Complex C and the IV Arch}
\shortauthors{Richter et al.}

\begin{document}

\title{The Diversity of High- and Intermediate-Velocity Clouds:\\ Complex C 
versus IV Arch\altaffilmark{1}}


\author{Philipp Richter\altaffilmark{2}, 
Kenneth R. Sembach\altaffilmark{3}, 
Bart P. Wakker\altaffilmark{2},
Blair. D. Savage\altaffilmark{2},
Todd M. Tripp\altaffilmark{4},
Edward M. Murphy\altaffilmark{3},
Peter M.W. Kalberla\altaffilmark{5}}

\and

\author{Edward B. Jenkins\altaffilmark{4}}


\altaffiltext{1}{Partly based on observations with the NASA/ESA Hubble Space
Telescope, obtained at the Space Telescope Science Institute, which
is operated by the Association of Universities for Research in Astronomy,
INC. under NASA contract NAS5-26555.}
\altaffiltext{2}{Department of Astronomy, University of Wisconsin-Madison,
475 N. Charter Street, Madison, WI\,53706; richter@astro.wisc.edu}
\altaffiltext{3}{Department of Physics and Astronomy, Johns Hopkins University,
3400 N. Charles Street, Baltimore MD\,21218}
\altaffiltext{4}{Princeton University Observatory, Peyton Hall, Princeton, 
NJ\,08544}
\altaffiltext{5}{Radioastronomisches Institut, Universit\"at Bonn, 53121 
Bonn, Germany}


\begin{abstract}

We present {\it Far Ultraviolet Spectroscopic Explorer} (FUSE) and {\it Space 
Telescope Imaging Spectrograph} (STIS) 
observations of interstellar ultraviolet absorption 
lines in the Galactic high-velocity cloud Complex C and the Intermediate 
Velocity
Arch (IV Arch) in direction of the quasar PG\,1259+593 ($l=120\fdg6$, 
$b=+58\fdg1$).
Absorption lines from C\,{\sc ii}, N\,{\sc i}, N\,{\sc ii},
O\,{\sc i}, Al\,{\sc ii}, Si\,{\sc ii}, P\,{\sc ii}, S\,{\sc ii}, Ar\,{\sc i}, 
Fe\,{\sc ii}, and Fe\,{\sc iii}
are used to study the atomic abundances in these two halo clouds
at $V_{\rm LSR} \sim -130$ km\,s$^{-1}$ (Complex C) and $-55$ 
km\,s$^{-1}$ (IV Arch).
The \ion{O}{1}/\ion{H}{1} ratio provides the best measure of the overall 
metallicity
in the diffuse interstellar medium, because ionization effects do not alter
the ratio, and oxygen is at most only lightly depleted from the gas into dust
grains.  For Complex C, we find an oxygen
abundance of 0.093 $^{+0.125}_{-0.047}$ solar,
consistent with the idea that Complex C represents the infall of low metallicity gas 
onto the Milky Way.  In contrast,
the oxygen abundance in the IV Arch is 0.98 $^{+1.21}_{-0.46}$ solar, 
which indicates a Galactic origin. 
We report the detection of an intermediate-velocity absorption component
at $+60$ km\,s$^{-1}$ that is not seen in \ion{H}{1} 21\,cm emission. 
The clouds along the PG\,1259+593 sight line have a variety of 
properties, proving that multiple processes are responsible
for the creation and circulation of intermediate- and high-velocity
gas in the Milky Way halo.
\end{abstract}




\section{Introduction}

Although significant progress has been made in the last few years
in exploring the distribution and chemical composition of intermediate
and high-velocity clouds (IVCs and HVCs, respectively) in the
halo of the Milky Way, an overall unified model for their formation
is still lacking. These clouds are typically defined as concentrations
of neutral hydrogen (H\,{\sc i}) at velocities inconsistent
with a simple model of differential Galactic rotation.
The distinction between IVCs and HVCs is loosely based on the
observed radial velocities of the clouds;
IVCs have radial velocities with respect to the 
Local Standard of Rest (LSR) of $30$ km\,s$^{-1} \leq |V_{\rm LSR}| 
\leq 90$ km\,s$^{-1}$, while
HVCs have velocities $|V_{\rm LSR}| > 90$ km\,s$^{-1}$.
Recent studies (Lu et al.\,1998; Wakker et al.\,1999; Gibson et al.\,2000, 
2001; Murphy et al.\,2000; Bluhm et al.\,2001; Sembach et al. 2001) reveal 
disparate chemical compositions for several of these clouds in different 
directions in the sky.
The kinematics and chemical make-up of the IVCs and some of the HVCs can be 
explained successfully by the ``Galactic fountain'' model
(Shapiro \& Field 1976; Houck \& Bregman 1990), in which hot gas ejected 
out of the Galactic disk by supernova explosions eventually
falls back in the form of condensed neutral
clouds moving at intermediate and high velocities. Clouds participating in this circulatory
pattern should have a nearly solar
metallicity reflecting that of their place of origin. 
There are, however, a number of HVCs whose abundances are inconsistent with 
the Galactic fountain model.  One such case is the Magellanic Stream, which has
abundances close to those of the Small Magellanic Cloud (SMC) 
and is believed to be tidally stripped 
out of the SMC system during a close encounter with
the Galaxy (e.g., Wannier \& Wrixon 1972; Lu et al.\,1998; Gibson et al.\,2000;
Sembach et al.\,2001).  High-velocity cloud Complex C has an even lower 
abundance ($\sim 0.1$ solar; Wakker et al.\,1999) that is inconsistent with 
gas originating in the disk of the Galaxy or in the Magellanic Clouds. 
Thus, Complex C might represent accreting metal-poor material from 
intergalactic space. Accretion of substantial quantities of metal-poor gas 
in the form of HVCs would have a significant influence on the chemical 
evolution of the Milky Way. 

It is clear that all of the
intermediate and high-velocity clouds in the Galactic halo do not have a 
single common origin.
Therefore, it is important to quantify the properties of IVCs and
HVCs to the greatest extent possible using ultraviolet (UV) absorption line 
spectroscopy. Many astrophysically important
atoms and molecules have their electronic transitions in
the wavelength region between 912 and 3000 \AA. 
Ultraviolet absorption line spectroscopy
is the most sensitive and accurate method to study
gas-phase abundances and physical properties
in diffuse interstellar clouds in the Galactic halo, where typical gas 
densities are significantly lower than in Galactic disk clouds.
The {\it Far Ultraviolet Spectroscopic Explorer} (FUSE) has sufficient 
sensitivity and spectral resolution to investigate absorption at 
$\lambda \le 1187$ \AA\ in the Galactic halo and beyond along
a large number of sight lines. At longer wavelengths ($\lambda > 
1150$ \AA), the {\it Space
Telescope Imaging Spectrograph} (STIS) on HST provides additional information 
for a number of atomic species and the Ly$\alpha$ absorption
line of neutral hydrogen near $1215.7$ \AA. Combining data from
these instruments provides a particulary powerful tool for 
investigating HVCs and IVCs.

In this study we use FUSE and STIS absorption line data to investigate the
intermediate- and high-velocity gas in the Galactic halo toward the 
quasar PG\,1259+593 ($V=15.84$; $z_{\rm em} = 0.478$; $l=120\fdg6$, $b=+58\fdg1$). 
This sight line passes through 
the Intermediate-Velocity Arch (IV Arch; Kuntz \& Danly 1996) and
the high-velocity cloud Complex C (see Wakker \& van\,Woerden 1997). 
Complex C has been the subject of several recent
absorption line studies along different lines of sight.  For example,
Wakker et al.\,(1999) studied the sulfur abundance of Complex C toward 
Mrk\,290 and found (S/H) $=0.089 \pm 0.024\,(^{+0.020}_{-0.005})$ solar
based on UV absorption line data in
combination with high resolution H\,{\sc i} data from the Westerbork
Radio Telescope and H$\alpha$ data from the {\it Wisconsin H$\alpha$ Mapper} 
(WHAM).   In contrast, Murphy et al\,.(2000) observed Mrk\,876 with FUSE
and found (\ion{Fe}{2}/\ion{H}{1}) $=0.48 \pm 0.2 (\pm 0.2)$ times the solar 
(Fe/H) ratio in Complex C, suggesting little or no depletion of Fe 
into dust grains.
Gibson et al.\,(2001) analyzed STIS data of the Complex C absorption along 
the sight lines
toward Mrk 279, Mrk 290, Mrk 501, Mrk 817, and Mrk 876, and found variations
in S\,{\sc ii}/H\,{\sc i} implying (S/H) in the range from 0.08 to 0.44 
solar. These variations could be due in part to ionization effects, although
intrinsic abundance variations within Complex C cannot be ruled out.

To avoid having to deal with substantial ionization corrections,
we have chosen to study the metal content in Complex C toward PG\,1259+593.
The sight line passes through Complex C in a region of relatively high 
H\,{\sc i} column density (($8.4 \pm 0.1) \times 10^{19}$ cm$^{-2}$; 
Wakker et al.\,2001), so ionization effects are much less important 
than at lower column densities. Other Complex C sight lines either have much
lower UV fluxes (e.g., Mrk 290) or substantially lower column densities
and greater confusion with lower velocity features (e.g., Mrk 279, Mrk 876).
In the direction of PG\,1259+593, the Complex C absorption is well-separated 
in velocity from the IV Arch, thus allowing a direct comparison of 
intermediate and high-velocity material in the Galactic halo 
{\it within the same spectrum}. The IV Arch lies at a height of $0.8-1.5$ kpc
above the Galactic plane (see Wakker 2001). An overall metallicity for the IV 
Arch has yet not been well determined, although several sulfur measurements 
(Spitzer \& Fitzpatrick 1993; Fabian et al., in preparation) suggest a 
near-solar metallicity, in striking contrast to the much lower metallicity of 
the more distant Complex C (d $>$ 6.1\,kpc; van\,Woerden et al.\,1999).
An extensive summary of absorption line measurements available for many
IVCs and HVCs is given by Wakker (2001).

This paper is organized as follows:
we review the sight line structure
toward PG\,1259+593 in \S2. In \S3 we present the FUSE and STIS observations 
of PG\,1259+593 and
our analysis methods. In \S4 we discuss absorption from gas associated
with Complex C; in \S5 we analyze absorption related to the IV Arch.
In \S6 we report the discovery of another IVC component at positive velocities.
\S7 provides a brief summary of the velocity distribution of 
the highly ionized species observed along the sight line.  Concluding remarks 
are given in \S8.

\section{The PG\,1259+593 Sight Line}

The H\,{\sc i} emission lines of the PG\,1259+593 sight line as observed with
the Effelsberg 100m telescope (Wakker et al.\,2001; $9\farcm1$ beam), 
the NRAO 140 foot telescope
at Green Bank (Murphy et al., in preparation; $21\farcm$0 beam),
and the Leiden-Dwingeloo Survey (Hartmann \& Burton 1997;
$36\farcm0$ beam) are shown in
the left panel of Figure\,1. Three major H\,{\sc i} components are visible at
$0, -55$ and $-130$ km\,s$^{-1}$, representing emission from
local Galactic gas, the IV Arch
close to core IV\,19 (Kuntz \& Danly 1996) in the lower Galactic halo, and
high-velocity cloud Complex C near the
CIII-C core (Giovanelli, Verschuur \& Cram 1973), respectively.
The local Galactic H\,{\sc i} emission component is weak because
the direction $l=120\fdg6$, $b=58\fdg1$ lies close to the ``Lockman Hole'', 
the region with the smallest amount of local H\,{\sc i} in the northern 
hemisphere (Lockman, Jahoda, \& McCammon 1986).
This makes the analysis of the intermediate- and high-velocity material
along this sight-line less complicated than in other directions because
1) the local component is not confused with absorption at intermediate 
velocities, and 2) local H$_2$ absorption, which can blanket a significant 
portion of the UV spectrum below $\sim1100$\,\AA, 
is weak.  There are substantial differences
in the three \ion{H}{1} emission profiles shown in Figure~1 resulting from 
the different instrumental resolutions of the data and beam-smearing effects.
We adopt the highest resolution (Effelsberg) data, yielding
N(\ion{H}{1}) = $(4.4 \pm 0.5) \times 10^{19}$ cm$^{-2}$ for the
local Galactic component at $0$ km\,s$^{-1}$,  
$(3.0 \pm 0.1) \times 10^{19}$ cm$^{-2}$ for the IV Arch at
$-55$ km\,s$^{-1}$, and $(8.4 \pm 0.1) \times 10^{19}$ cm$^{-2}$
for Complex C at $-130$ km\,s$^{-1}$ (Wakker et al.\,2001).
\footnote{The H\,{\sc i} column densities derived from the lower resolution 
data are $(4.9 \pm 1.9) \times 10^{19}$ cm\,$^{-2}$ (Green Bank; GB) and 
$(3.6 \pm 0.5) \times 10^{19}$ cm\,$^{-2}$ (Leiden-Dwingeloo; LD) for
the local Galactic component,
$(3.6 \pm 1.5) \times 10^{19}$ cm\,$^{-2}$ (GB) and
$(2.9 \pm 0.4) \times 10^{19}$ cm\,$^{-2}$ (LD) for
the IV Arch, and
$(6.3 \pm 1.6) \times 10^{19}$ cm\,$^{-2}$ (GB) and
$(4.1 \pm 0.4) \times 10^{19}$ cm\,$^{-2}$ (LD) for
Complex C.}

PG\,1259+593 has a redshift of $z_{\rm em}=0.478$ (Marziani et al.\,1996), and 
there are several intergalactic H\,{\sc i} Ly\,$\alpha$ absorbers 
along this sight line 
(e.g., Bahcall et al.\,1993). Intergalactic absorption detected
in the FUSE and STIS data will be discussed in a future paper. For the present
study, it is sufficient to note that these IGM features do not affect the 
analysis of the absorption associated with Complex C and the IV Arch 
presented in this paper.

\newpage

\section{Observations and Data Reduction}

FUSE is equipped with four co-aligned optical channels (2 SiC channels for
905-1105 \AA\, and 2 LiF channels for 1000-1187 \AA) and two microchannel
plate detectors.  
A detailed description of the FUSE instrument and its on-orbit
performance is given by Moos et al.\,(2000) and Sahnow et al.\,(2000).
The FUSE observations of PG\,1259+593 were obtained on 25
February 2000, 25 December 2000, and 29 January 2001 through the 
large apertures (LWRS, $30\arcsec\times30\arcsec$) of the four channels and 
were recorded in photon address mode.  
Sixty-four exposures totaling $\sim 193$ ks.
are available for all four channels, although 
the light of the quasar was not optimally centered in all of the 
apertures throughout the observations and the detector high voltage was 
down during part of one observation.  Table 1 provides an overview of the 
FUSE observations of PG\,1259+593.  A detailed analysis of the raw time-tagged
photon list was required to eliminate small event bursts of unknown origin 
from the raw data (see Sahnow et al.\,2000). 

The data were extracted using the 
standard FUSE calibration pipeline CALFUSE (v1.8.7) available at the Johns 
Hopkins University, and the individual exposures for each channel were coadded
using a cross-correlation technique to remove small relative wavelength 
shifts between exposures.  The pipeline corrects for geometrical distortions 
in the detector grid, the Doppler shifts introduced by the motions of Earth 
and satellite through space, and small spectral shifts due to 
thermally-induced motions of the diffraction gratings.  This version of the
pipeline does not correct for fixed-pattern noise or the spatial curvature 
of the lines introduced by the astigmatism of the optical system, which 
degrades the spectral resolution slightly at most wavelengths.
To improve the signal-to-noise
ratio (S/N), the data were rebinned over 5 pixels, resulting in
$\sim 10$ km\,s$^{-1}$ wide bins. No additional smoothing was applied.
The velocity resolution of the combined FUSE data is
$\sim 25$ km\,s$^{-1}$ (FWHM).
The average flux in the spectrum is 
$\sim 2 \times 10^{-14}$ erg\,cm$^{-2}$\,s$^{-1}$\,\AA$^{-1}$, equivalent to
an average  S/N of $\sim 10$ per (rebinned) pixel element.

STIS observations of PG1259+593 were carried out on 17--19 January 2001 with
the intermediate-resolution far-UV echelle grating (E140M) and the 
$0\farcs 2 \times 0\farcs 06$ slit.  Thirty-four exposures
totaling 81 ksec were obtained. In this observing mode, STIS
provides a resolution of 7 km s$^{-1}$ (FWHM) and
wavelength coverage from $\sim$1150 to 1730 \AA\ with only five small
gaps between orders at $\lambda >$ 1634 \AA. 
We reduced the data with
the STIS Investigation Definition Team (IDT) version of CALSTIS at the
Goddard Space Flight Center. The individual spectra were flatfielded,
extracted, and wavelength and flux calibrated with the standard
procedures. The STIS IDT correction for scattered light was then applied,
and the individual spectra were co-added weighted by their inverse
variances averaged over a large, high signal-to-noise region.  Finally,
overlapping regions of adjacent orders were co-added with weighting based
on inverse variance. For further information on the design and on-orbit
performance of STIS, see Woodgate et al.\,(1998) and Kimble et al.\,(1998).

The FUSE and STIS absorption profiles were continuum-normalized by fitting
low-order polynomials to the data; a selection of normalized profiles
is shown in Figure.\,1.  (For display purposes only, we have rebinned the 
STIS data in this figure to $\sim 10$ km\,s$^{-1}$ wide pixel elements, 
similar to the FUSE data.)  We measured equivalent widths for the weakly 
ionized species in the HVC and IVC gas by fitting single Gaussian profiles 
to the main absorption components at $-130$ km\,s$^{-1}$ (Complex C),
$-55$ km\,s$^{-1}$ (IV Arch), and for the IVC at $+60$ km\,s$^{-1}$.
For every line in the FUSE bandpass, equivalent widths were measured 
independently using data for 
the two highest S/N channels, and 
then the values were averaged with weights proportional to the inverse squares
of the respective errors.
Wavelengths, oscillator strengths (Morton 1991; 2001, in preparation),
and equivalent widths (and 1$\sigma$ errors) 
for all measured lines are listed in Table 2.
The cited errors include contributions from photon statistics and
continuum placement uncertainties.
For the determination of column densities
we made use of a standard curve-of-growth technique.

\section{Complex C}

Metal line absorption from weakly ionized species associated with
Complex C is detected near $-130$ km\,s$^{-1}$ 
in C\,{\sc ii}, C\,{\sc iii},
N\,{\sc i}, O\,{\sc i}, Al\,{\sc ii}, Si\,{\sc ii}, S\,{\sc ii}, 
and Fe\,{\sc ii}.
Weak lines from other species (e.g., Ar\,{\sc i} $\lambda1048.220$ 
and P\,{\sc ii} 
$\lambda 1152.818$) are not detected at Complex C velocities, but $3\sigma$ 
upper limits were derived (see Table 2).
Measurements of the equivalent widths of C\,{\sc ii}, C\,{\sc iii},
 and Si\,{\sc iii} 
in Complex C are not possible since these strong
lines extend over the entire velocity range between
$-180$ to $+180$ km\,s$^{-1}$ (see Figure\,1).
The strong lines of oxygen, silicon, and
iron exhibit an asymmetry in the red wing of their absorption
profiles, implying the presence of a  weak absorption component near
$-110$ km\,s$^{-1}$. Multiple Complex C components have also been
observed toward other background sources (e.g., Mrk\,876, see Gibson
et al.\,2001), where H\,{\sc i} emission and absorption by weakly ionized
species arises in at least two distinct components between $-100$ and 
$-200$ km\,s$^{-1}$. 
The present PG\,1259+593 data do not allow us to analyze the two 
absorptions separately for all species, so we will consider the effect of 
this weak component on our column density determinations by examining its 
effect on the \ion{O}{1} and \ion{Si}{2} lines observed by STIS.
The present Effelsberg H\,{\sc i} 21cm emission data do not give any guidance
as to the existence of 
multiple H\,{\sc i} emission components, but it is possible to
include a narrow, weak (log $N$(\ion{H}{1})$<18.5$) emission component
at $\sim -110$ km\,s$^{-1}$ 
without a noticeable change in the observed profile.

All of the Complex C lines from weakly ionized species collectively fit 
a single-component curve of growth with 
$b=9.8^{+4.7}_{-1.2}$ km\,s$^{-1}$ (Figure\,2), suggesting
that the velocity-dispersion parameter is determined by turbulent motions. 
There is a systematic trend for
the stronger lines (e.g., O\,{\sc i} $\lambda 1302.169$) to favor
a higher $b$-value than the weak lines, supporting the idea that
in this direction Complex C consists of two or more unresolved subcomponents.
Column densities and gas-phase abundances relative to the solar
values (Anders \& Grevesse 1989; Grevesse \& Noels 1993) were
derived by fitting the data to a single-component curve of growth with $b=9.8$ 
km\,s$^{-1}$; they are
presented in Table 3 along with their respective $1\sigma$ errors. 

To estimate the influence of a possible unresolved subcomponent 
at $\sim -110$ km\,s$^{-1}$ on the Complex C column densities, 
we have created composite apparent column density profiles 
(see Savage \& Sembach 1991) for the
strong absorption lines of oxygen and silicon in the STIS data. 
A weak component at  $-114$ km\,s$^{-1}$ is clearly visible.
Simulating the influence of this 
component on the overall Complex C line profile, we find
that such a component modifies the curve of growth of the
main component in a very gentle manner, and the resulting curve is very
similar to the curve of growth with $b=9.8$ km\,s$^{-1}$,
derived from our single-component Gaussian fitting analysis.
We estimate an uncertainty of $\approx 0.05$ dex (12\%) 
for the abundances in Complex~C due to the existence
of this unresolved absorption component (note that this uncertainty is not 
included in the formal errors cited in Table 3). 
The presence of this subcomponent does not significantly alter the conclusions
of our abundance study of Complex C.

All elements considered in this study have normalized gas-phase abundances 
defined by
[$X$/H] $\equiv$ log($N_X/N_{\rm H\,{\sc I}})- $log($X$/H)$_{\odot}< -0.8$ dex,
except phosphorus, for which only a $3\sigma$ upper limit of
$\le -0.27$ could be derived. 
Comments regarding the limits for Ar\,{\sc i} and Fe\,{\sc ii}
are found in subsequent paragraphs.
The normalized interstellar gas-phase abundances
for all measured species in Complex C are shown in Figure\,3.
An accurate measure for the mean metallicity in the ISM
is the O\,{\sc i}/H\,{\sc i} ratio; oxygen is not significantly depleted 
onto
dust grains (Meyer et al.\,1998) and the O\,{\sc i}/H\,{\sc i} ratio is not
altered by ionization effects (i.e., (\ion{O}{1}/\ion{H}{1}) $\approx$ (O/H)). 
This is because O\,{\sc i} and H\,{\sc i}
have similar ionization potentials and both elements are strongly
coupled through charge exchange interactions (e.g., Sofia \& Jenkins 1998). 
For Complex C,
we find [O/H]$=-1.03^{+0.37}_{-0.31}$, equivalent to $0.093^{+0.125}_{-0.047}$ 
times the solar (O/H) ratio.
The value agrees very well with the low sulfur abundance found
toward Mrk 290 (Wakker et al.\,1999), but is
significantly lower than the sulfur and iron abundances derived from 
S\,{\sc ii} and Fe\,{\sc ii}
toward Mrk 279, Mrk 501, Mrk 817, and Mrk 876 (Gibson et al.\,2001; Murphy 
et al.\,2000).
For S\,{\sc ii} we find [S/H]$=-0.85^{+0.12}_{-0.15}$. Thus, the inferred 
normalized gas-phase abundance of sulfur is significantly higher than for 
oxygen, but S\,{\sc ii} 
exists in regions were hydrogen is both neutral and ionized. Assuming that
the sulfur-to-oxygen ratio in Complex C is the same as in local Galactic gas
and that the residual difference in the measured values is caused by 
S\,{\sc ii} in ionized gas along the sight line,
we estimate an average degree of ionization of H$^+$/(H$^0$+H$^+)=0.32$ and
$N$(H$_{\rm total})=N($H$^0)+N($H$^+)=1.26 \times 10^{20}$ cm$^{-2}$.
\footnote{We note that the S/O ratio in Complex C might differ from the solar
    value because of a different nucleosynthetic history. If so,
    the ionization fraction would be different.}

If UV radiation is the major source of ionization in Complex~C, it would be 
straight forward to explain the non-detection of neutral argon because 
\ion{Ar}{1}
has a large photoionization cross section (Sofia \& Jenkins 1998).  Comparing
\ion{Ar}{1} to \ion{H}{1}, we find [Ar/H]$\le -1.42$,
but it is possible that the actual value for [Ar/H] is significantly
{\it higher}, if most of the Ar is in the form of Ar\,{\sc ii}.
For similar reasons, a significant fraction of the nitrogen may also be 
ionized, which could account for the very low normalized abundance based on 
\ion{N}{1} ([N/H]$=-1.94^{+0.17}_{-0.21}$). The N\,{\sc ii}
$\lambda 1083.994$ absorption profile observable with FUSE is consistent 
with a strength of $\sim 130$ m\AA\ required to correct the the total 
nitrogen abundance up to 0.1 solar.  However, this line 
has low S/N and it is blended by Fe\,{\sc ii}
$\lambda 1083.420$. We derive an upper limit of $183$ m\AA\,($3\sigma$)
for N\,{\sc ii} absorption at Complex C velocities.

Although the normalized gas-phase abundances derived for O, Al, Si, S, and 
Fe are consistent with a single value to within the errors,
the data show weak evidence for a slight underabundance of iron
in comparison to the other elements.
We find [Si/H]$=-0.91^{+0.28}_{-0.27}$ and
[Al/H]$=-0.98^{+0.30}_{-0.50}$, and [Fe/H]$=-1.27^{+0.20}_{-0.14}$,
based on the data for Si\,{\sc ii}, Al\,{\sc ii}, and Fe\,{\sc ii}.
The abundance of Al is derived from a single
saturated line (\ion{Al}{2} 
$\lambda 1670.787$) and therefore
has a relatively large error bar.
In the local interstellar medium, Si, Al, and Fe 
are all depleted into dust grains and their
gas-phase abundance is significantly lower than that
of mildly depleted species, such as O (e.g., Savage \& Sembach 1996).
For Complex C, however, we find [Si/H]$\approx$
[Al/H]$\approx$[O/H], indicating
that silicon and aluminum are probably not incorporated
into dust grains in Complex C if the ionization corrections for 
\ion{Si}{2} and \ion{Al}{2} are modest. 
While the ionization potentials for Si\,{\sc ii}
and Fe\,{\sc ii} are almost identical, the photoionization 
cross section of Fe\,{\sc ii} is much higher than that of Si\,{\sc ii}
once the ionizing photons exceed an energy of $\sim 20$ eV
(see Figure\,3 in Sofia \& Jenkins 1998). Thus, its appears plausible
that a significant fraction of the missing iron is in Fe\,{\sc iii} because
of ionizing radiation at high energies.

We can check on the amounts of doubly ionized species in Complex~C.
Si\,{\sc iii}  $\lambda 1206.500$
absorption is present at Complex C velocities in the STIS spectrum, but 
this line cannot be used to derive a Si\,{\sc iii} column density because the
three velocity components are heavily overlapping (see Figure\,1).
The Fe\,{\sc iii} $\lambda 1122.524$ line is observed by FUSE but it
is blended by Galactic Fe\,{\sc ii} $\lambda 1121.975$ (Figure\,1). 
Estimating the strength
of the blending Fe\,{\sc ii} $\lambda 1121.975$ absorption by comparing with
Fe\,{\sc ii} $\lambda 1096.877$ (the two lines have  similar values of
log $f\lambda$ -- see Table\,2) yields a Galactic Fe\,{\sc ii} component 
strength of $\approx 50$ m\AA. The equivalent width
of Galactic Fe\,{\sc ii} $\lambda 1121.975$ plus
Fe\,{\sc iii} $\lambda 1122.524$ at $-130$ km\,s$^{-1}$
has a very similar strength,
suggesting that there is no significant
Fe\,{\sc iii} absorption at Complex~C velocities
($<33$ m\AA\ ($3 \sigma$)). 
We derive log $N$(Fe\,{\sc iii})$\le 13.67$ and 
[(Fe\,{\sc ii}$+$Fe\,{\sc iii})/H\,{\sc i}]
$= -1.15$, or 0.24 dex lower than 
[Si/H] based solely on \ion{Si}{2}. 
Therefore,
the apparently low abundance
of iron in Complex C is not likely to result from ionization effects.
More likely, a low iron abundance in Complex C could be due to depletion of iron
into Fe-rich dust grains. As discussed by Savage \& Sembach (1996), dust-phase abundance
measurements of Galactic halo clouds indicate the presence of dust cores
that contain Mg and Fe but do not contain much Si.
An alternative, and perhaps more speculative hypothesis, would be that
a low iron abundance in Complex~C reflects an overabundance of 
$\alpha$ elements ({\rm S, Si}, and {\rm O}) compared to iron.
An enhanced abundance of $\alpha$ elements in Complex C has also been 
suggested for the Mrk 876, Mrk 290, and Mrk 817 sight lines based upon
comparisons of N\,{\sc i}, S\,{\sc ii}, and Fe\,{\sc ii} column densities
(Gibson et al.\,2001; Murphy et al.\,2000). 
However, considering that the abundances of iron and other elements are consistent {\it within
their error range}, more accurate data is required to investigate the apparent
iron deficiency in Complex C on a statistically more reliable basis.

Summarizing, we find that Complex C has an overall abundance of $\sim$0.1 
solar in the direction of PG\,1259+593, strongly supporting the idea that 
Complex C has an origin outside the Milky Way.
Additional high quality data for other sight lines may help to quantify the 
abundance pattern in Complex C and determine whether the apparent abundance 
variations seen in Complex C by Gibson et al. (2001) and the 
apparent iron deficiency toward PG\,1259+593 are intrinsic in nature or due 
to a combination of effects (e.g., nucleosynthetic history, ionization, dust).

\section{IV Arch}

Absorption lines of weakly ionized metals C\,{\sc ii}, C\,{\sc iii},
N\,{\sc i}, O\,{\sc i}, Al\,{\sc ii}, Si\,{\sc ii}, Si\,{\sc iii}, 
S\,{\sc ii}, and Fe\,{\sc ii} are detected at the velocity of the IV Arch 
($-55$ km\,s$^{-1}$).  Absorption lines of Ar\,{\sc i} and P\,{\sc ii} are
not detected.  The equivalent widths are listed in Table 2.
The data fit best on a curve of growth
with $b=10.2^{+5.7}_{-3.2}$ km\,s$^{-1}$ (Figure\,2), implying
macroscopic turbulence and/or unresolved sub-components.
Column densities and gas-phase abundances, based on the
single component fit to the curve of growth with
$b=10.2$ km\,s$^{-1}$, are presented in Table 3.
Figure\,3 provides an overview of normalized interstellar gas-phase
abundances in the IV Arch (see also \S4).
The oxygen abundance in the IV Arch is [O/H]$=-0.01^{+0.35}_{-0.27}$, or
$0.98^{+1.21}_{-0.46}$ solar, 
in striking contrast to
the low oxygen abundance for Complex C derived in the previous 
section. The sulfur abundance (derived from S\,{\sc ii}) is slightly higher, 
[S/H]$=+0.11^{+0.11}_{-0.08}$, suggesting
an ionization fraction of H$^+$/(H$^0$+H$^+)=0.24$ using the same line 
of reasoning in \S4.
Similar to Complex C,  argon and nitrogen are
very underabundant in comparison to oxygen and sulfur
([Ar/H]$\le -0.98$ and [N/H]$=-0.68^{+0.50}_{-0.29}$,
based on the data for Ar\,{\sc i} and N\,{\sc i}).
These low abundances are most likely due to photoionization since neither 
argon nor nitrogen is readily incorporated into dust. 
Low argon and nitrogen abundances have been seen also 
in other intermediate-velocity clouds
(e.g., Richter et al.\,2001).
Abundances for silicon (Si\,{\sc ii})
and iron (Fe\,{\sc ii}) are [Fe/H] = $-0.22^{+0.28}_{-0.33}$ dex and [Si/H] = 
$-0.26^{+0.19}_{-0.15}$,
suggesting relatively mild depletions into dust grains.

The nearly solar abundances and the $z$-height bracket of $0.8-1.5$ kpc
suggest that the IV Arch 
has its origin in the Milky Way disk, for instance
as part of a Galactic fountain (Shapiro \& Field 1976; 
Houck \& Bregman 1990).
The abundances derived here agree well with 
those for the LLIV Arch toward PG\,0804+761 (Richter et al.\,2001). 
The FUSE and STIS data of PG\,1259+593 therefore suggest that the IV 
Arch and Complex C have completely different origins.

\section{IVC at +60 km\,s$^{-1}$}

The strong absorption lines of O\,{\sc i}, C\,{\sc ii}, C\,{\sc iii}, 
Al\,{\sc ii}, Si\,{\sc ii}, Si\,{\sc iii}, and Fe\,{\sc ii} reveal
the presence of another IVC component at {\it positive} velocities
near +60 km\,s$^{-1}$. Absorption at positive velocities is not
expected for interstellar gas participating in Galactic rotation in this 
direction. 
Equivalent widths for this absorption component 
are listed in Table 2. 
Neither the Effelsberg nor the Green Bank H\,{\sc i} data
show emission at $+60$ km\,s$^{-1}$, suggesting that the H\,{\sc i}
column density is below the detection limit. Another 
possibility is that the diameter of this cloud is very small
and that beam-smearing effects make this component undetectable
in H\,{\sc i} emission. 
The Effelsberg data yield a $3 \sigma$ upper limit of
$N$(H\,{\sc i})$\le 3.0 \times 10^{18}$ cm$^{-2}$ for this IVC.
H\,{\sc i} absorption in the higher
Lyman lines (Ly\,$\gamma$ and higher) may be present, but
blending with the Galactic H\,{\sc i} component and
other lines in combination with low S/N at these wavelengths
prevents any useful quantitative analysis of the  H\,{\sc i} 
absorption. Thus, it is impossible to determine abundances
for this IVC with the present data.

\section{Highly Ionized Species}

In addition to the weakly ionized species,
absorption lines of  
C\,{\sc iv} $\lambda 1548.195$, Si\,{\sc iv} $\lambda 1393.755$, and
O\,{\sc vi} $\lambda  1031.926$ are seen in both Complex C and the IV Arch.
This is the second detection of O\,{\sc vi} in Complex C 
(see Murphy et al.\,2000
for the profiles toward Mrk\,876). The O\,{\sc vi}
absorption profile is shown in Figure.\,1.
O\,{\sc vi} absorption at negative velocities might also be present in
the other O\,{\sc vi} line at $\lambda 1037.617$, but
is blended by Galactic H$_2$\,$(5-0)$\,R(1).
The high ion absorption at Complex C velocities suggests the presence of 
a highly ionized boundary of Complex C caused by
interaction with surrounding material in the Galactic halo
(e.g., Sembach et al.\,2000; Murphy et al.\,2000).

\section{Concluding Remarks}

The FUSE spectrum of PG\,1259+593 shows the diverse
nature of intermediate- and high-velocity clouds in the halo of
the Milky Way. Absorption line data obtained so far suggest that
IVCs generally tend to have abundances similar to those in the
disk of the Milky Way (for a summary see Wakker 2001) . 
Abundances found in HVCs, in contrast, are not explainable by
a single origin. Metallicities of $\sim 0.1$ solar, as found for Complex C 
in this study and others (e.g., Wakker et al.\,1999),
indicate that metal-deficient gas is falling into the Milky Way halo.
Other high-velocity clouds, such as the one in front of the 
LMC (Bluhm et al.\,2001, Richter et al.\, in preparation),
have significantly
higher metal abundances, which can be explained by 
the Galactic fountain model.
The Magellanic Stream (Lu et al.\,1998; Gibson et al.\,2000;
Sembach et al.\,2001) has abundances similar to that of the SMC
and probably is the result of tidal interaction between 
the Milky Way and the SMC.

High-velocity cloud Complex C remains one of the most
interesting cases of Galactic high-velocity gas investigated so far. 
Our [O/H] abundance for Complex C agrees well with the [S/H] abundance
found by Wakker et al.\,(1999) toward Mrk\,290, but studies of 
other sight lines
suggest that there might be significant abundance variations in Complex~C 
(Gibson et al.\,2001; Murphy et al.\,2000). The data obtained so far 
suggests that there might be
a slight over-abundance of $\alpha$ elements in comparison to iron in Complex~C.
With this study, 
an abundance pattern including eight elements 
(N, O, Al, Si, P, S, Ar, and Fe) is now available for comparison with other
sight lines. Further investigations of Complex C, in particular the study of 
its deuterium and molecular hydrogen content, may help to understand the 
abundances derived in this paper.

\newpage

\acknowledgments

This work is based on data obtained for the
the Guaranteed Time Team by the NASA-CNES-CSA FUSE
mission operated by the Johns Hopkins University.
Financial support has been provided by NASA
contract NAS5-32985. Further support for this work was provided
by NASA through grant number GO-08695.01-A from the
Space Telescope Science Institute.  KRS acknowledges partial support
from NASA Long Term Space Astrophysics grant NAG5-3485. We thank
J.C. Howk and W.P. Blair for helpful comments on this work.

\clearpage
\newpage
\includegraphics{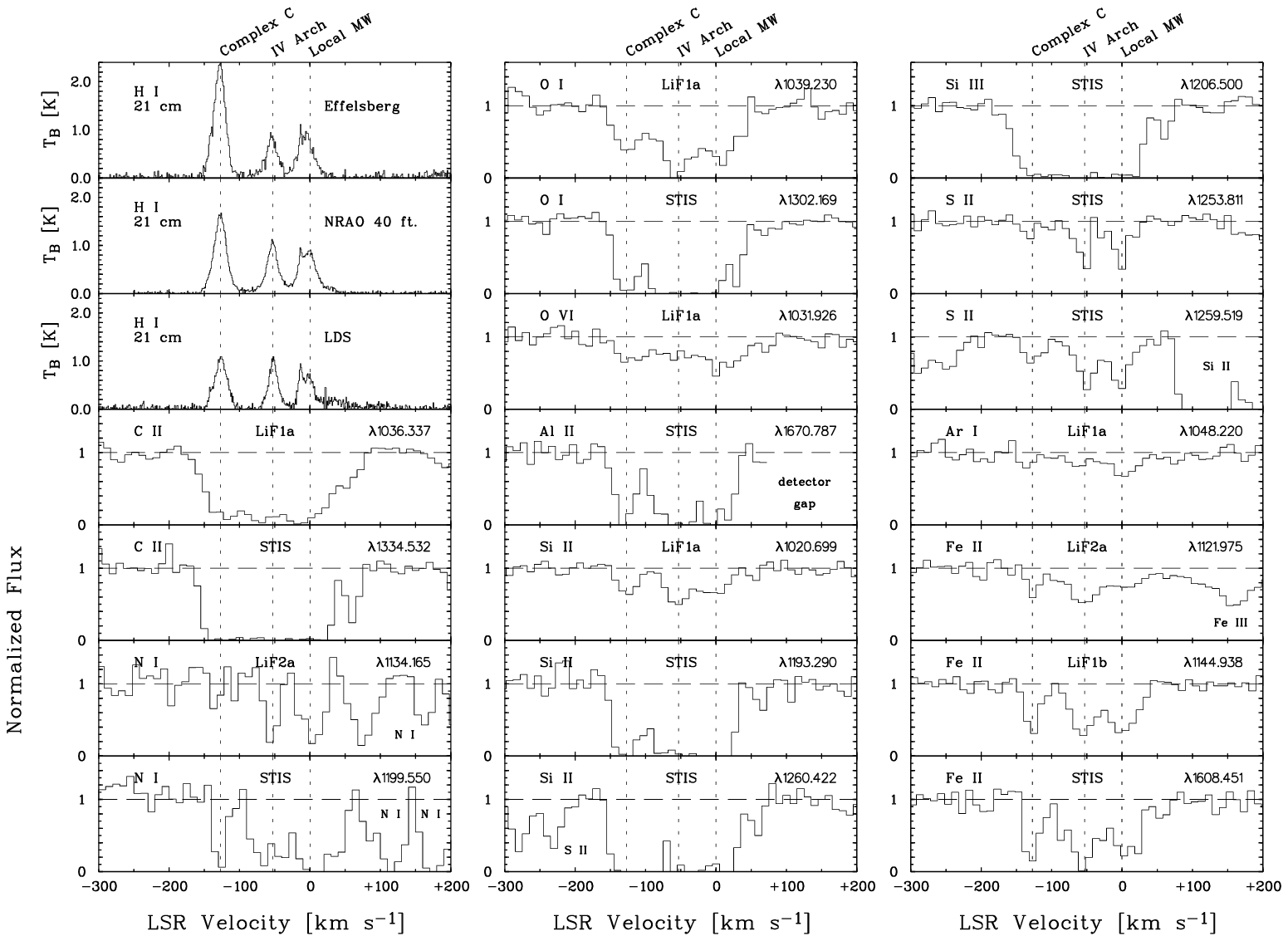}
\figcaption[pg1259fig01.ps]{
Continuum-normalized absorption profiles of several species
from FUSE and STIS data of PG\,1259+593, plotted against 
LSR velocity. 
For this plot, the data have been rebinned to 10 km\,s$^{-1}$ wide pixels.
The three major absorption components (local Galactic gas at
0 km\,s$^{-1}$, IV Arch at $-55$ km\,s$^{-1}$, and Complex C at 
$-130$ km\,s$^{-1}$) are marked with dotted lines and are identified 
above the boxes. Absorption at $+60$ km\,s$^{-1}$ is visible only in the
strong lines of C, O, Al, Si, and Fe. 
The three H\,{\sc i} emission profiles 
(Effelsberg data with a $9\farcm1$ beam, NRAO 140 ft. data
with a  $21\farcm0$ beam, and
Leiden-Dwingeloo data with a $36\farcm0$ beam)
show substantial differences, indicating
beam smearing effects. 
\label{fig1}}

\clearpage
\newpage
\includegraphics{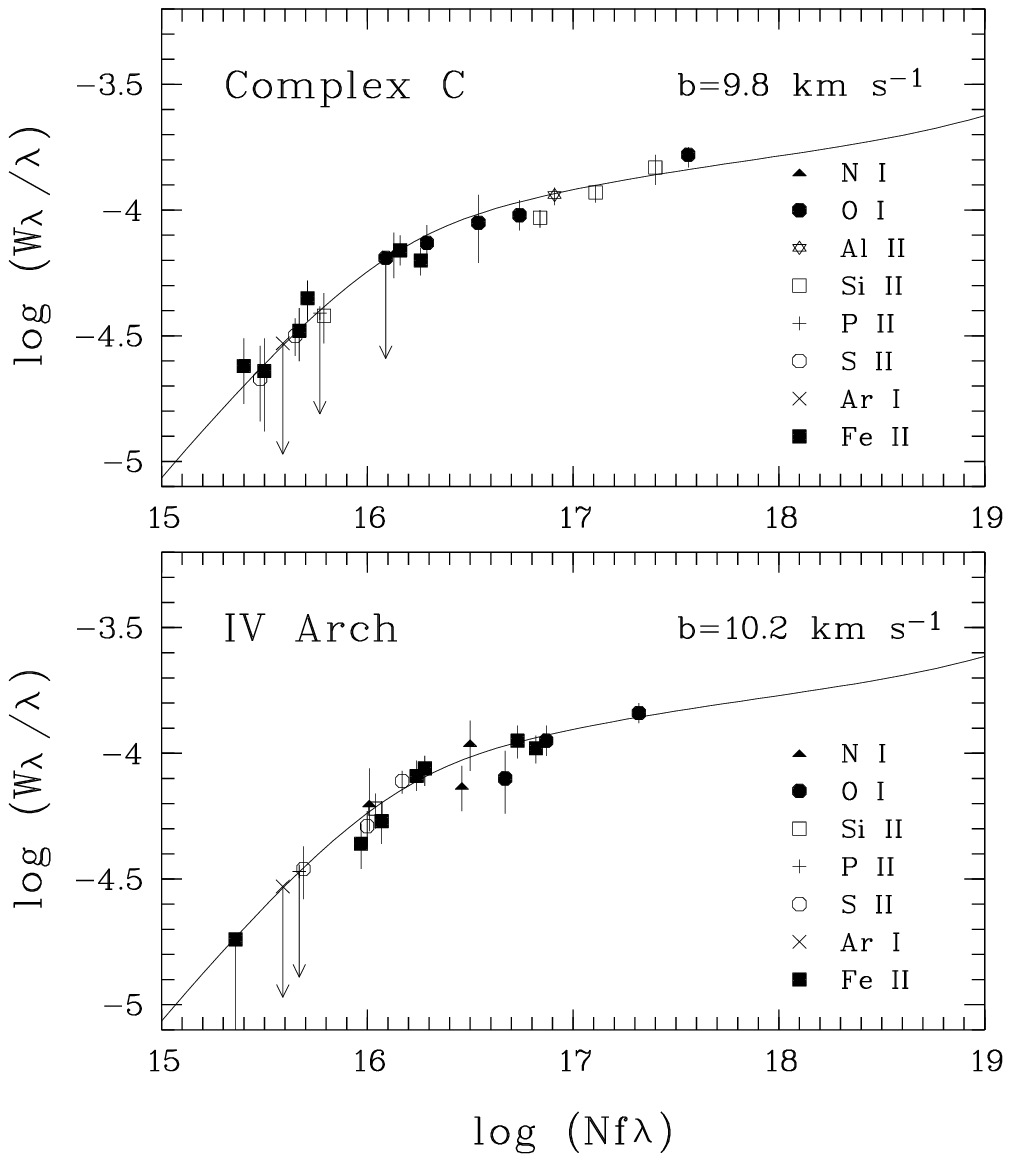}
\figcaption[pg1259fig02.ps]{
Empirical curves of growth for ions observed in Complex C (upper 
panel) and 
the IV Arch (lower panel). The different ions are labeled at the lower right
side of each panel.
For Complex C, the data points fit best on a curve
of growth with $b=9.8^{+4.7}_{-1.2}$ km\,s$^{-1}$. For the IV Arch, the best 
fit is found for $b=10.2^{+5.7}_{-3.2}$ km\,s$^{-1}$. 
\label{fig2}}

\clearpage
\newpage
\includegraphics{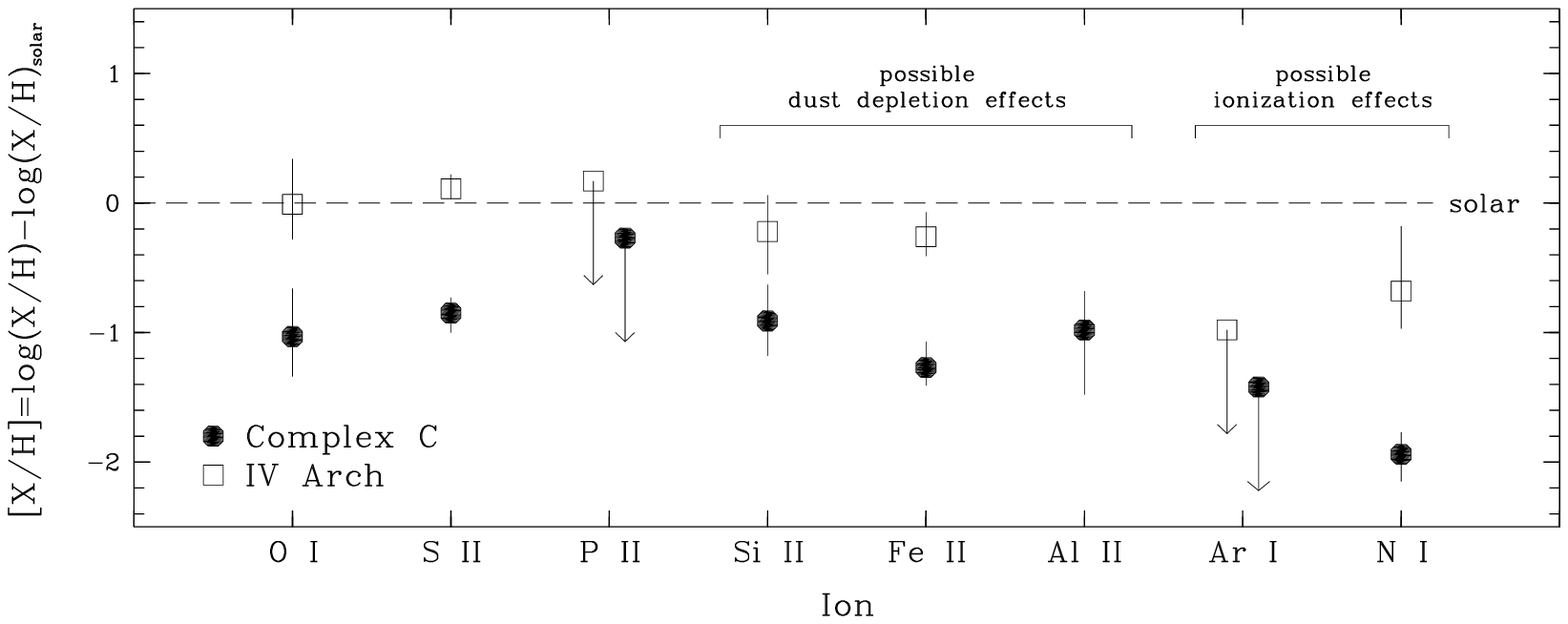}
\figcaption[pg1259fig03.ps]{
Normalized interstellar gas-phase abundances in Complex C and the IV Arch.
Abundances in Complex C are systematically lower than in the IV Arch,
suggesting a different enrichment history of both clouds.
Ionization and dust depletion effects are discussed in \S4 and \S5.
Error bars represent uncertainties from both equivalent-width
measurement errors (see \S3) and errors from the fits to the curves of growth. 
\label{fig3}}

\clearpage
\begin{deluxetable}{ccccccc}
\tabletypesize{\normalsize}
\tablecaption{Log of PG\,1259+593 FUSE and STIS Observations}
\tablewidth{0pt}
\tablehead{
\colhead{Instrument} & \colhead{Dataset ID} & \colhead{Obs. Date} & \colhead{Aperture} & \colhead{Exp. Time} & \colhead{Number of Exp.} & \colhead{Notes}\\
\colhead{} & \colhead{} & \colhead{} & \colhead{} & \colhead{[ks]}}
\startdata
FUSE & P1080101 & 2000 Feb.\,25  &  $30\arcsec\times30\arcsec$  & 52.4 & 12 & 1\\
FUSE & P1080102 & 2000 Dec.\,25  &  $30\arcsec\times30\arcsec$  & 57.9 & 21 & 2 \\
FUSE & P1080103 & 2001 Jan.\,29  &  $30\arcsec\times30\arcsec$  & 82.2 & 31 & 3\\
\\
HST STIS & 8695 & 2001 Jan.\,17-19 & $0\farcs2 \times 0\farcs06$   & 81.3 & 34 &\\
\enddata
\tablenotetext{\,}{Notes:}
\tablenotetext{\,}{1) No SiC\,2 channel data. SiC\,1 and LiF\,2 channels were not optimally centered at all times.} 
\tablenotetext{\,}{2) LiF\,1 data only. SiC\,1, SiC\,2, and LiF\,2 channels were not co-aligned - no flux recorded.}
\tablenotetext{\,}{3) Detector 1 shut down during exposures 17-29.}
\end{deluxetable}

\clearpage
\newpage
\begin{deluxetable}{lrcrrrl}
\tabletypesize{\normalsize}
\tablecaption{Atomic Absorption Lines Associated with Complex C, the IV Arch,
and the IVC at $+60$ km\,s$^{-1}$ \label{tbl-2}}
\tablewidth{0pt}
\tablehead{
\colhead{Species} & \colhead{$\lambda_{\rm vac}$\,$^{\rm a}$}   & \colhead{log\,
$\lambda f^{\rm a}$}   &
\colhead{$W_{\lambda}$\,$_{\rm Complex C}$$^{\rm b}$} & \colhead{$W_{\lambda}$\,
$_{\rm IV Arch}$$^{\rm b}$}  &
\colhead{$W_{\lambda}$\,$_{\rm IV+60}$$^{\rm b}$} & \colhead{Instrument}\\
\colhead{} & \colhead{[\AA]}   & \colhead{}   &
\colhead{[m\AA]} & \colhead{[m\AA]} & \colhead{[m\AA]} & 
}
\startdata
C\,{\sc ii} \dotfill  & 1036.337 & 2.104 & \nodata     & \nodata & $\le$ 80 & FUSE LiF1a, LiF2b\\
                      & 1334.532 & 2.232 & \nodata     & \nodata & 72 $\pm$ 9 & HST STIS\\
N\,{\sc i} \dotfill   & 1134.165 & 1.238 & $\le$ 41    & 72  $\pm$ 17 & \nodata & FUSE LiF2a\\ 
                      & 1134.980 & 1.693 & \nodata     & 84 $\pm$ 17  & \nodata & FUSE LiF2a\\
                      & 1199.550 & 2.202 & 81 $\pm$ 16 & \nodata\     & \nodata & HST STIS\\
                      & 1200.710 & 1.725 & \nodata     & 132 $\pm$30  & \nodata & HST STIS\\
N\,{\sc ii} \dotfill  & 1083.994 & 2.097 & $\le$ 183   & \nodata      & \nodata & FUSE SiC1a,SiC2b\\
O\,{\sc i} \dotfill   &  929.517 & 0.329 & $\le$ 48    & 74 $\pm$ 22  & \nodata & FUSE SiC1b, SiC2a\\
                      &  936.630 & 0.534 & 70 $\pm$ 11 & 106 $\pm$ 14 & \nodata & FUSE SiC1b,SiC2a\\
                      &  948.686 & 0.778 & 84 $\pm$ 25 & \nodata      & \nodata & FUSE SiC2a\\
                      & 1039.230 & 0.980 & 100 $\pm$ 13 & 151  $\pm$ 15 & \nodata & FUSE LiF1a, LiF2b\\
                      & 1302.169 & 1.804 & 214 $\pm$ 20 & \nodata       & 22 $\pm$ 7 & HST STIS\\
Al\,{\sc ii} \dotfill & 1670.787 & 3.486 & 186 $\pm$ 15 & \nodata       & $\le$33 & HST STIS\\
Si\,{\sc ii} \dotfill & 1020.699 & 1.225 &  39 $\pm$ 9 & 61  $\pm$ 10   & \nodata & FUSE LiF1a\\
                      & 1190.416 & 2.543 & \nodata     & \nodata        & 24 $\pm$ 7 & HST STIS\\
                      & 1193.290 & 2.844 & 175 $\pm$ 25 & \nodata       & 29 $\pm$ 6 & HST STIS\\
                      & 1260.422 & 3.104 & \nodata      & \nodata       & 40 $\pm$ 7 & HST STIS\\
                      & 1304.370 & 2.284 & 121 $\pm$ 9 &  \nodata       & $\le$ 16 & HST STIS\\  
                      & 1526.707 & 2.546 & 178 $\pm$ 13 & \nodata       & \nodata & HST STIS\\
Si\,{\sc iii} \dotfill & 1206.500 & 3.304 &  \nodata    & \nodata       & 36 $\pm$ 6 & HST STIS\\
P\,{\sc ii} \dotfill  & 1152.818 & 2.451 &  $\le$ 37   & $\le$ 37  & \nodata     & FUSE LiF1b, LiF2a\\
S\,{\sc ii} \dotfill  & 1250.584 & 0.834 &  $\le$ 19   & 43 $\pm$ 10    & \nodata & HST STIS\\
                      & 1253.811 & 1.135 & 24 $\pm$ 8 & 64 $\pm$ 10     & \nodata & HST STIS\\
                      & 1259.519 & 1.311 & 40 $\pm$ 7 & 97 $\pm$ 9      & \nodata & HST STIS \\  
Ar\,{\sc i} \dotfill  & 1048.220 & 2.440 &  $\le$ 28    & $\le$ 28      & \nodata & FUSE LiF1a, LiF2b\\
Fe\,{\sc ii} \dotfill & 1096.877 & 1.554 & 49 $\pm$ 9 & 95 $\pm$ 13     & \nodata & FUSE LiF2a\\
                      & 1121.975 & 1.512 &  37 $\pm$ 9 & 92  $\pm$ 13   & \nodata & FUSE LiF1b, LiF2a\\
                      & 1125.448 & 1.244 & 27 $\pm$ 8 & 49 $\pm$ 10     & \nodata & FUSE LiF2a\\
                      & 1142.366 & 0.633 & $\le$ 21   & $\le$ 21        & \nodata & FUSE LiF1b, LiF2a\\
                      & 1143.226 & 1.342 & 26 $\pm$ 9 & 62 $\pm$ 12     & \nodata & FUSE LiF2a\\ 
                      & 1144.938 & 2.096 & 72 $\pm$ 9 & 120  $\pm$ 16   & \nodata & FUSE LiF1b, LiF2a\\
                      & 1608.451 & 1.998 & 112 $\pm$ 16 & 182 $\pm$ 29 & 26 $\pm$ 6 & HST STIS\\
Fe\,{\sc iii} \dotfill & 1122.524 & 2.260 & $\le$ 33  & $\le$ 47 & \nodata & FUSE LiF2a, LiF1b\\
\enddata
\tablenotetext{a}{Vacuum wavelengths and oscillator strengths from Morton (1991; 2001, in preparation).}
\tablenotetext{b}{Equivalent widths and $1 \sigma$ errors (or $3 \sigma$ upper limits) are listed
for Complex C, the IV Arch, and the IVC at $+60$ km\,s$^{-1}$ (denoted IV$+60$).}
\end{deluxetable}

\clearpage
\newpage
\begin{deluxetable}{lrcrrrr}
\tabletypesize{\normalsize}
\tablecaption{Atomic Column Densities and Gas-Phase Abundances in Complex C and IV Arch \label{tbl-3}}
\tablewidth{0pt}
\tablehead{
\colhead{Species} & \colhead{I.P.} & \colhead{log($X$/H)$_{\odot}$\,$^{\rm a}$} 
&  \colhead{log $N_{\rm Complex C}$} &
\colhead{[$X$/H]$_{\rm Complex C}$\,$^{\rm b}$} & \colhead{log $N_{\rm IV Arch}$} 
& \colhead{[$X$/H]$_{\rm IV Arch}$\,$^{\rm b}$}\\
\colhead{} & \colhead{[eV]} & \colhead{+12} & \colhead{$N$ in [cm$^{-2}$]} & \colhead{}
 & \colhead{$N$ in [cm$^{-2}$]}}
\startdata
H\,{\sc i}   & 13.60 & 12.00 & 19.92 $\pm$ 0.01$^{\rm c}$       & \nodata & 19.48 $\pm$ 0.01$^{\rm c}$ & \nodata \\
C\,{\sc ii}  & 24.38 & 8.55  & \nodata          & \nodata       & \nodata          & \nodata \\
N\,{\sc i}   & 14.53 & 7.97  & 13.95$^{+0.17}_{-0.21}$     & $-$1.94$^{+0.17}_{-0.21}$  & 14.77$^{+0.50}_{-0.29}$ & $-$0.68$^{+0.50}_{-0.29}$ \\
N\,{\sc ii}   & 29.60 & 7.97  & $\le$15.71 &  \nodata       & \nodata          & \nodata \\
O\,{\sc i}   & 13.62 & 8.87  & 15.77$^{+0.37}_{-0.31}$ & $-$1.03$^{+0.37}_{-0.31}$ & 16.34$^{+0.35}_{-0.27}$ & $-$0.01$^{+0.35}_{-0.27}$ \\
Al\,{\sc ii} & 18.82 & 6.48    & 13.42$^{+0.30}_{-0.50}$ & $-$0.98$^{+0.30}_{-0.50}$ & \nodata & \nodata \\
Si\,{\sc ii} & 16.35 & 7.55  & 14.56$^{+0.28}_{-0.27}$ & $-$0.91$^{+0.28}_{-0.27}$   & 14.81$^{+0.28}_{-0.33}$ & $-$0.22$^{+0.28}_{-0.33}$ \\
P\,{\sc ii}  & 19.73 & 5.57  & $\le$ 13.22 & $\le$ $-$0.27        & $\le$ 13.22      & $\le$ $+$0.17 \\
S\,{\sc ii}  & 23.33  & 7.27  & 14.34$^{+0.12}_{-0.15}$  & $-$0.85$^{+0.12}_{-0.15}$ & 14.86$^{+0.11}_{-0.08}$ & $+$0.11$^{+0.11}_{-0.08}$ \\
Ar\,{\sc i}  & 15.76 & 6.65  & $\le$ 13.15  & $\le$ $-$1.42      & $\le$ 13.15      & $\le$ $-$0.98 \\
Fe\,{\sc ii} & 16.16 & 7.51  & 14.16$^{+0.20}_{-0.14}$ & $-$1.27$^{+0.20}_{-0.14}$  & 14.73$^{+0.19}_{-0.15}$ & $-$0.26$^{+0.19}_{-0.15}$ \\
Fe\,{\sc iii} & 30.65 & 7.51  & $\le$13.67 & $\le$ $-$1.76  & $\le$13.86 & $\le -$1.13\\

\enddata
\tablenotetext{a}{Anders \& Grevesse (1989); Grevesse \& Noels (1993).}
\tablenotetext{b}{[$X$/H] = log($N_X/N_{\rm H\,{\sc I}})- $log($X$/H)$_{\odot}$, limits are $3\sigma$.
Errors do not include an additional 12\% uncertainty due to possible
sub-structure within the Complex C absorption profiles.}
\tablenotetext{c}{From Effelsberg 21\,cm emission line data with a beamsize
of $9\farcm1$ centered on PG\,1259+593.}
\end{deluxetable}

\end{document}